\documentclass[proof]{pasj00}

\begin{document}
\SetRunningHead{Tosaki et al.}{Molecular Gas in M33}

\title{NRO M33 All Disk Survey of Giant Molecular Clouds (NRO MAGiC):
I. H{\sc i} to H$_2$ Transition}



%
 \author{%
   Tomoka \textsc{Tosaki}\altaffilmark{1},
   Nario \textsc{Kuno}\altaffilmark{2},
   Sachiko \textsc{Onodera}\altaffilmark{2},
    Rie \textsc{Miura}\altaffilmark{3},
   Tsyuyoshi \textsc{Sawada}\altaffilmark{4,5}\\
   Kazuyuki \textsc{Muraoka}\altaffilmark{6},
   Kouichiro \textsc{Nakanishi}\altaffilmark{4},
   Shinya \textsc{Komugi}\altaffilmark{4,5},
   Hiroyuki \textsc{Nakanishi}\altaffilmark{7}\\
   Hiroyuki \textsc{Kaneko}\altaffilmark{8},
   Akihiko \textsc{Hirota}\altaffilmark{2},
   Kotaro \textsc{Kohno}\altaffilmark{9,10}
   and
   Ryohei \textsc{Kawabe}\altaffilmark{2}}
 \altaffiltext{1}{Joetsu University of Education, Yamayashiki-machi, Joetsu, Niigata 943-8512, Japan}
 \email{tosaki@juen.ac.jp}
 \altaffiltext{2}{Nobeyama Radio Observatory, National Astronomical Observatory of Japan,\\
 Nobeyama, Minamimaki, Minamisaku, Nagano 384-1305, Japan}
  \altaffiltext{3}{Department of Astronomy, School of Science, The University of Tokyo\\
  The University of Tokyo, Hongo, Bunkyo, Tokyo 113-0033, Japan}
 \altaffiltext{4}{National Astronomical Observatory of Japan, Osawa, Mitaka, Tokyo 181-8588, Japan}
  \altaffiltext{4}{Joint ALMA Office, Alonso de C¥'{o}rdova 3107, Vitacura, Santiago 763-0355, Chile}
 \altaffiltext{6}{Department of Physical Science, Osaka Prefecture University, Gakuen 1-1, Sakai, Osaka 599-8531, Japan}
 \altaffiltext{7}{Graduate School of Science and Engineering, Kagoshima University, 1-21-35 Korimoto, \\
 Kagoshima, Kagoshima 890-0065, Japan}
 \altaffiltext{8}{The Graduate University for Advanced Studies (Sokendai), Osawa, Mitaka, Tokyo 181-8588, Japan}
  \altaffiltext{9}{Institute of Astronomy, School of Science, The University of Tokyo,  \\
  Osawa, Mitaka, Tokyo 181-0015, Japan}
\altaffiltext{10}{Research Center for Early Universe, School of Science, The University of Tokyo, \\
Hongo, Bunkyo, Tokyo 113-0033, Japan}

\KeyWords{galaxies: individual (M33) -- galaxies: ISM -- radio lines: galaxies} 

\maketitle

\begin{abstract}

We present the results of the Nobeyama Radio Observatory (NRO) M33 All Disk ($30^\prime \times 30^\prime$, or 7.3 kpc $\times$ 7.3 kpc) Survey of
Giant Molecular Clouds (NRO MAGiC) based on $^{12}$CO ($J=1\mbox{--}0$) observations using the NRO 45-m telescope.
The spatial resolution of the resultant map is 19$^{\prime\prime}$.3,
corresponding to 81 pc, which is sufficient to identify each Giant Molecular Cloud (GMC)
in the disk.
We found clumpy structures with a typical spatial scale of $\sim$100 pc,
corresponding to GMCs, and no diffuse, smoothly distributed component
of molecular gas at this sensitivity.
The overall distribution of molecular gas roughly agrees with
that of H{\sc i}. However, closer inspection of the CO and H{\sc i} maps
suggests that not every CO emission is associated with local H{\sc i} peaks,
particularly in the inner portion of the disk ($r <$ 2 kpc), although
most of CO emission is located at the local H{\sc i} peaks in the outer
radii.
We found that most uncovered GMCs are accompanied by
massive star-forming regions, although the star formation rates (SFRs)
vary widely from cloud to cloud.
The azimuthally averaged H{\sc i} gas surface density exhibits a flat
radial distribution. However, the CO radial distribution shows
a significant enhancement within the central 1--2 kpc region,
which is very similar to that of the SFR.
We obtained a map of the molecular fraction,
$f_{\rm mol} = \Sigma_{\rm H_2}/(\Sigma_{\rm H{\sc i}}+\Sigma_{\rm H_2})$, at a 100-pc resolution.
This is the first $f_{\rm mol}$ map covering an entire galaxy with a GMC-scale resolution.
We find that $f_{\rm mol}$ tends to be high near the center.
The correlation between $f_{\rm mol}$ and gas surface density
shows two distinct sequences.
The presence of two correlation sequences can be explained by differences in metallicity, i.e., higher ($\sim$2-fold) metallicity in the central region ($r<$ 1.5 kpc) than in the outer
parts. Alternatively, differences in
scale height can also account for the two sequences, i.e.,
increased scale height toward the outer disk.

\end{abstract}

\section{Introduction}

The interstellar medium (ISM) is one of the essential components of galaxies
because stars are born in and return to them.
The ISM is greatly influenced by massive stars due to their stellar winds
and supernova after their death. Therefore, the formation process of
massive stars within the ISM is crucial for understanding
of the nature and evolution of galaxies.
In the Milky Way Galaxy, a large fraction of the ISM is in the form of Giant
Molecular Clouds (GMCs; \cite{sand1985,sco1987}),
and they are known to be major sites of massive star formation (\cite{wal1987}).

At this spatial scale, i.e., a few tens to a hundred pc
we can find a wide variety of star-formation activities and the nature of molecular clouds.
For example, in the LMC, the variation in star-formation activity
from cloud to cloud also becomes visible at this spatial scale,
and it is interpreted as indicating different evolutionary stages of GMCs
along the star formation processes (\cite{Kawa2009}).
Furthermore, no tight correlation between the molecular gas masses and star formation rates (SFRs)
can be found among GMCs in M33 seen at a scale of $\sim$100 pc or smaller (\cite{ono2010}),
despite the fact that there exists a tight correlation between the molecular gas masses and SFRs
at scales of a few thousand to a few hundred pc in galaxies, known as the Kenicutt--Schmidt law (K-S law; Kennicutt 1998).
The reported breakdown of the K-S law in M33 is presumably due to
the differences in the star-formation properties, i.e., GMCs at different evolutionary stages
(\cite{ono2010}).
With $\sim$100-pc-scale observations of multitransition CO lines,
the coexistence of GMCs in various evolutionary stages has
also been reported within the supergiant H{\sc ii} region NGC 604 in M33, suggesting
that sequential star formation is on-going (\cite{Tosa2007,miur2010}).
Consequently, a survey of GMCs at a spatial scale of $\sim$100 pc
is essential for understanding of the processes of massive
stars from the ISM.

In addition to this, an unbiased survey is also important
because it can sample GMCs in various environments,
i.e., in the central region, outer region, spiral arms,
inter-arm regions, etc. Formations of GMCs,
dense molecular gas, and massive stars will be affected
not only by local processes within individual GMCs
but also by these global structures of a galaxy
or environments within a galaxy.

Furthermore, the transition from atomic gas to molecular gas is also
essential to understanding the nature and
evolution of the ISM,
since the first phase of star formation process is the formation of a dense gas
cloud from the diffuse ISM.
Many studies of the transition from atomic to molecular gas have been conducted
using both theoretical and observational approaches.

\citet{elm1993} introduced the molecular fraction $f_{\rm mol}$
as a parameter to quantify the transition.
It is the ratio of the molecular gas surface density to the total gas surface
density including both atomic and molecular gases.
$f_{\rm mol}$ can be expressed as a function of metallicity, UV radiation
field, and gas pressure.
In particular, metallicity is considered to have a strong impact
on the transition of a gas from atomic to molecular.
Observationally, the presence of a ``molecular front,'' i.e., a sharp transition
front from molecular to atomic, has been identified in some nearby
galaxies (\cite{hon1995, sof1995,nak2006a}).
This suggests that the transition from atomic gas to molecular gas
occurs within a very narrow span along the galactic radius,
and the variation in interstellar metallicity has been suggested
to play a crucial role in this phase transition
(\cite{hon1995}).
However, these previous studies on the gas phase transition
were based on observations at a resolution of a few hundred pc.
It is important to understand the transformation process
of the gas phase at the GMC scale, an essential spatial scale for the ISM,
in order to obtain a unified picture of the change in the ISM from the diffuse atomic phase
to massive stars.

M33 is one of the nearest spiral galaxies in the local group ($D$ = 840 kpc; \cite{free1991}).
The proximity of M33 allows us to resolve the individual GMCs with
the existing large-aperture single dishes.
Since M33 has a relatively small inclination angle of 51$^{\circ}$ (\cite{deul1987}),
we can obtain a comprehensive view of a galaxy.
This is a key advantage for the study of the correlation between GMC
properties and galactic structures such as spiral arms.

There are many previous studies on the ISM and star
formation in M33.
Whole-disk surveys of molecular gas in the $^{12}$CO ($J=1\mbox{--}0$) line
have been conducted with the BIMA interferometer (\cite{eng2003}) and FCRAO 14-m telescope (\cite{hey2004}),
and \citet{ros2007} combined the results of these data with observations with
the 45-m telescope at the Nobeyama Radio Observatory (NRO)
toward a part of the disk.
A partial mapping of $^{12}$CO ($J=2\mbox{--}1$) has also been conducted with
the IRAM 30-m telescope and a comparison between the CO and H{\sc i} distributions has been made (\cite{Grat2010}).
Selected regions of M33, such as the supergiant H{\sc ii} region NGC 604,
have also been mapped in $^{12}$CO ($J=1\mbox{--}0$) lines
using the OVRO array (\cite{wil1989}, \yearcite{wil1990}) and NMA (\cite{miur2010}),
and $^{12}$CO ($J=1\mbox{--}0$)/$^{12}$CO ($J=3\mbox{--}2$) multitransition maps of NGC 604 have been presented
using the NRO 45-m and ASTE 10-m dishes (\cite{Tosa2007}).
Atomic gas (H{\sc i}) in M33 has been studied with the VLA and Effesberg 100-m
telescope (\cite{deul1987}), and the cold dust component
has been mapped in the 1.1 mm (AzTEC/ASTE; \cite{kom2011})
and Herschel SPIRE bands (250--500 $\mu$m; \cite{kram2010}),
showing a decline in cold dust temperature as a function of radius.

Here we present the initial results of the NRO M33 All Disk Survey of Giant Molecular Clouds (NRO MAGiC) project
conducted using the NRO 45-m telescope.
By using the largest millimeter-wave telescope equipped with a multibeam
receiver system (\cite{Sun2000}) combined with an efficient on-the-fly (OTF)
technique (\cite{saw2008}), we have obtained a sensitive map
of $^{12}$CO ($J=1\mbox{--}0$) emission over almost the entire region of M33
with a uniform quality of spectra at a GMC-scale spatial resolution.
The survey area, sensitivity, and spatial resolution of this map is compared with that of others
of M33 in Table \ref{tab:1}.

Initial results of the survey (northern half of the map)
have already been used to show the breakdown of the K-S law
at $<$100-pc scales (\cite{ono2010}), emphasizing the importance
of $\sim$100-pc-scale observations of the ISM and star formation.

In this paper, we present an overview of the whole-disk $^{12}$CO ($J=1\mbox{--}0$) survey of M33
using the NRO 45-m telescope, with particular focus on the atomic-to-molecular gas
phase transition at the GMC scale.

\begin{table}
  \caption{Wide-area CO	observatons of M33}\label{tab:1}
 \begin{center}
 \begin{tabular} {llcccc}
\hline	
 transition & telescope & area & \multicolumn{2}{c}{resolution} & noise level  \\
&  & &  spatial & velocity  &    \\
&  & (arcmin$^2$) & (arcsec) & (km s$^{-1}$) & (mK)  \\
 \hline
$J=1\mbox{--}0$ & BIMA\footnotemark[$*$]       &1100 & 13    & 2  &  240 \\
& FCRAO 14 m\footnotemark[$\dagger$]                                 & 960 & 50     &  1  &  53   \\
& BIMIA + FCRAO 14 m + NRO 45 m\footnotemark[$\ddagger$]    & 172 & 20     &  2.6 & 60  \\
& NRO 45 m\footnotemark[$\S$]                                      & 900 &19    & 2.5  & 130  \\
\\
$J=2\mbox{--}1$ & IRAM 30 m\footnotemark[$\|$]   & 650 & 12 & 2.6 & 20--50 	\\
\hline
 \multicolumn{4}{@{}l@{}}{\hbox to 0pt{\parbox{85mm}{\footnotesize
 \par\noindent
 \footnotemark[$*$] \cite{eng2003}
 \par\noindent
 \footnotemark[$\dagger$] \cite{hey2004}
 \par\noindent
 \footnotemark[$\ddagger$]  \cite{ros2007} 
 \par\noindent
 \footnotemark[$\S$]  this work
\par\noindent
 \footnotemark[$\|$]  \cite{Grat2010}	 
  }\hss}}
    \end{tabular}
  \end{center}
\end{table}
\section{Observations}

\begin{figure*}
  \begin{center}
    \FigureFile(80mm,80mm){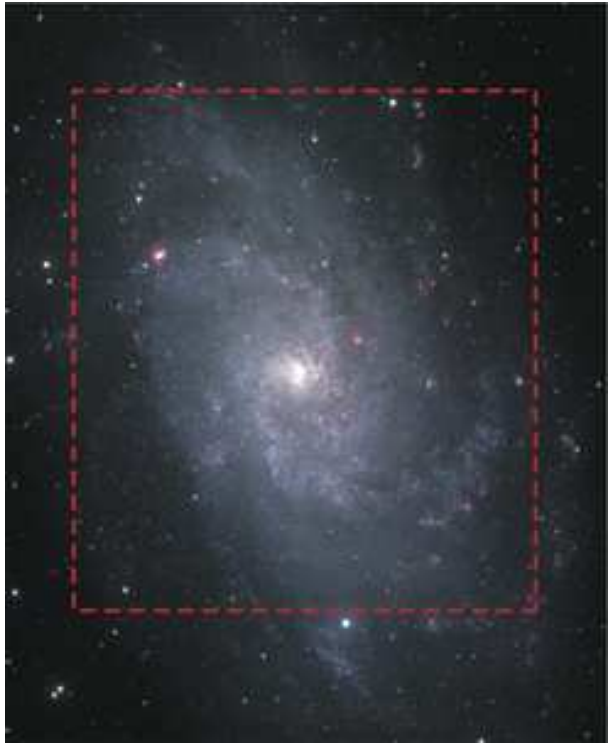}
  \end{center}
  \caption{Observed area (dashed box, $30^\prime \times 30^\prime$) superposed on a three-color composite optical image of M33 obtained with the Subaru telescope: 
  courtesy of V. Vansevicius, S. Okamoto, and N. Arimoto.
  The optical image consists  of three colors,  blue  (B), green (V), and red (H$\alpha$).  The H$\alpha$ image is not continuum-subtracted.}\label{fig:1}
\end{figure*}

We observed the $^{12}$CO ($J=1\mbox{--}0$) emission toward the disk of M33.
Figure~\ref{fig:1} shows our observed area, $30 ^\prime \times 30^\prime$, corresponding to 7.3 kpc $\times$ 7.3 kpc.
It covers most of the molecular gas disk and includes many star forming regions.

The observations were conducted from January 2008 to April 2009 using
the NRO 45-m telescope equipped with 5 $\times$ 5 focal-plane SIS array receivers (BEARS)
capable of simultaneously observing 25 positions in the sky
(\cite{Sun2000}).
Using a 100-GHz SIS receiver (S100) equipped with a single sideband filter, NGC 7538 was observed
to measure the sideband ratio of each beam of BEARS, which consists of double-side-band receivers.
The main beam efficiency measured with S100 was $\eta_{\rm mb}$ = 0.32 $\pm$ 0.02 at 115 GHz and
the errors in the scaling factors of BEARS were smaller than $\sim$20\%.
Wide-band digital autocorrelators (AC) were used to cover a velocity width of 1332 km s$^{-1}$
with a velocity resolution of 2.6 km s$^{-1}$ at 115.271204 GHz.
We monitored a SiO maser source IRC+30021 with the five-points observation method using a 43-GHz SIS receiver (S40)
every 1 h to check the pointing accuracy.
Note that the final map was made using data with pointing errors smaller than 7.5$^{\prime\prime}$ and
with wind speeds lower than 10 m s$^{-1}$ to avoid the systematic intensity loss due to the pointing errors.
The resultant total observation time was 136 h.

The observations were performed with the OTF mapping technique (\cite{saw2008}).
The ``scanning noise'' was removed by combining the horizontal and perpendicular scans using basket weaving.
The data reductions were made with the OTF reduction software package NOSTAR, which was implemented by the NRO.
Although the beam size of the telescope was 15$^{\prime\prime}$,
the spatial resolution of the final map was 19.3$^{\prime\prime}$ due to the grid spacing of 7.5$^{\prime\prime}$.

The rms noise levels in the velocity channel map with a velocity range of 2.6 km s$^{-1}$ and
the total integrated intensity map were 130 mK and 1.6 K km s$^{-1}$, respectively, in the main beam temperature scale.
The noise level in the integrated intensity map corresponds to 9.6 $\times 10^4 M_\odot$ in the beam size (80 pc)
by applying the Galactic CO-to-H$_2$ conversion factor, 3 $\times 10^ {20}$ cm$^{-2}$ (K km s$^{-1}$)$^{-1}$ (\cite{wil1990}).
This value corresponds to a limiting mass surface density of 15$M_\odot$pc$^{-2}$, including a correction of He, 
and is much lower than the surface density of GMCs in the Milky Way, $40\mbox{--}100 M_\odot$pc$^{-2}$ (\cite{hey2009}). 

\section{Results}

\subsection{Total integrated intensity map and velocity field}

\begin{figure*}
  \begin{center}
    \FigureFile(150mm,80mm){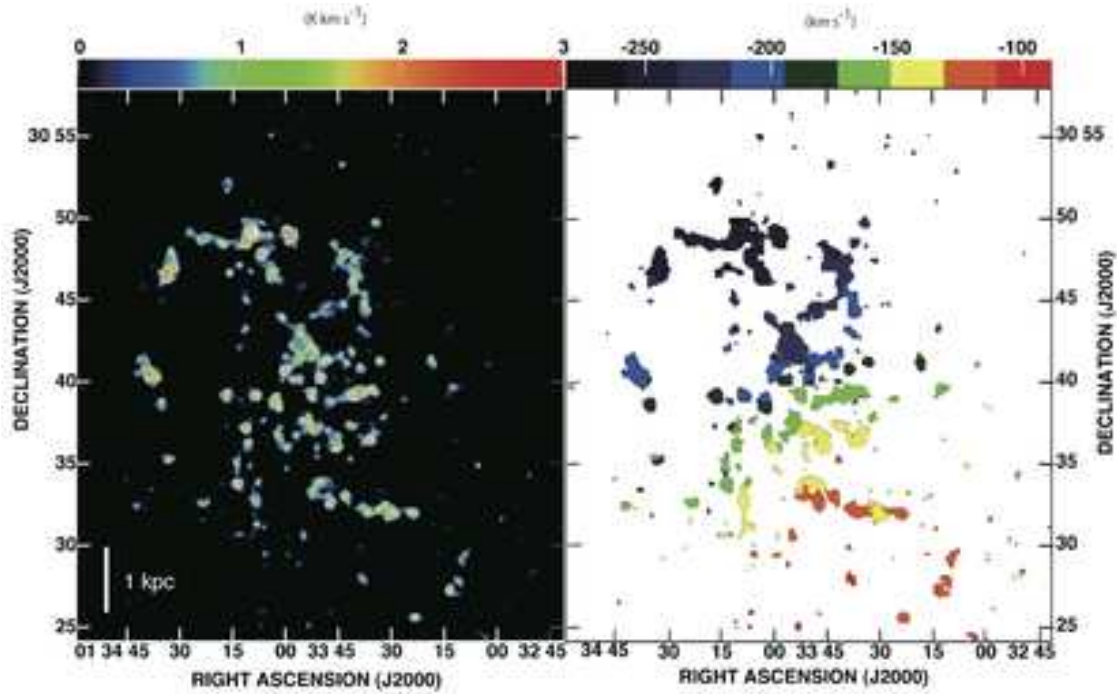}
  \end{center}
  \caption{Total integrated intensity map (left) and intensity-weighted mean radial velocity map (right) of $^{12}$CO ($J=1\mbox{--}0$) emission
   acquired with the NRO 45-m telescope.
   The map size is $30' \times 30'$, and the effective spatial resolution is $19''$,
   corresponding to $7.3$ kpc $\times 7.3$ kpc and 81 pc at the distance of M33 ($D$ = 840 kpc), respectively.
The contour interval is 0.5 K km s$^{-1}$ in the $T_a^*$ scale, corresponding to 1 $\sigma$.}\label{fig:2}
\end{figure*}

Figure~\ref{fig:2} shows the total integrated intensity map and intensity-weighted mean radial velocity map of $^{12}$CO ($J=1\mbox{--}0$) emission.
This figure reveals many clumps with a typical size of $\sim$100 pc,
similar to that of GMCs in the Milky Way (\cite{sco1987}) rather than that of Giant Molecular Associations
 (GMAs; \cite{ran1990}).
There are chains of such "GMCs" in the northern and southern spiral arms.
Note that most of the emission consists of such GMCs and no diffuse emission is seen in the map 
above the limiting mas surface density of 15 $M_\odot$pc$^{-2}$.
We see no concentration of CO at the galactic center in this map, unlike in most molecule-rich galaxies (\cite{Saka1999}).
These properties are consistent with maps obtained by previous studies (\cite{ros2007}).

The velocity field of the $^{12}$CO ($J=1\mbox{--}0$) emission in figure~\ref{fig:2} is similar to that of H{\sc i} emission (\cite{deul1987}),
except for a few GMCs (e.g., GMCs located in the southern arm).
The velocity field is globally dominated by galactic rotation, and shows no clear evidence for streaming motion along spiral arms,
unlike grand design spiral galaxies (e.g., M51; \cite{kuno1997}).
This is consistent with low amplitude of arm strength in M33 (\cite{reg1994}).

\subsection{Comparison with atomic gas and SFR}

\begin{figure*}
  \begin{center}
    \FigureFile(150mm,80mm){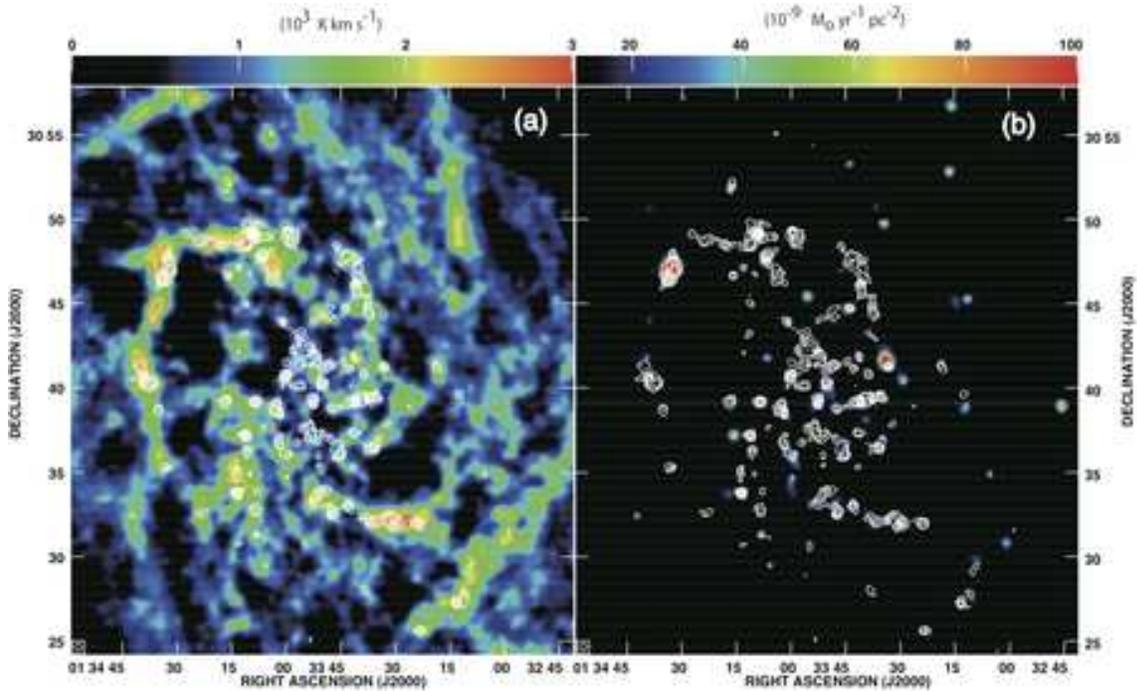}
  \end{center}
  \caption{Total integrated intensity map of $^{12}$CO ($J=1\mbox{--}0$) emission (contours) superposed on H{\sc i} emission (left, \cite{ros2007}) and
  star-formation rate (right) derived from H$\alpha$ luminosity (\cite{hoo2000})
  with extinction-correction using MIPS 24-$\mu$m data (\cite{rie2004}).
  The contour interva and lowest contour aew same as figure~\ref{fig:2}. }\label{fig:3}
\end{figure*}

Figure~\ref{fig:3} shows CO intensity maps of M33 superposed on the H{\sc i} emission (left) and
SFR (right).
The H{\sc i} data of M33 are from observations made with the Westerbork Synthesis Radio Telescope (WSRT) combined those made with the Effelsberg 100-m telescope
 in order to compensate for the missing short spacings of the WSRT interferometry data (\cite{deul1987}).
The original data, with a resolution of $12^{\prime\prime} \times 24^{\prime\prime}$, was convolved
to a resolution of $24^{\prime\prime} \times 24^{\prime\prime}$ (\cite{ros2007}).

 The H{\sc i} emission map shows a more wide-spread distribution than that of the CO emission, and
 we found CO emission where H{\sc i} emission was distributed.
However, we also found different tendencies in the correlation between the CO and H{\sc i} distributions
for the outer and inner regions of M33:
most CO clouds are associated with local H{\sc i} peaks in the outer regions,
 whereas no such clear correspondence can be found in the inner regions.

The SFR was calculated using the relation between the SFR
and the extinction-corrected H$\alpha$ line emission presented by \citet{cal2007}.
The H$\alpha$ image of M33 was used by \citet{hoo2000}.
The Multiband Imaging Photometer for Spitzer (MIPS) 24-$\mu$m data retrieved by the Spitzer Science Center (\cite{rie2004}) were used for extinction-correction. 
The detailed data reduction process for the 24-$\mu$m images is explained by \citet{ono2010}.
The calculated SFR is shown on the right in figure~\ref{fig:3}(b) as a color scale together with the CO emission (contour).
This figure shows that most of the GMCs are associated with star-forming regions.
However, we also find star-forming regions without GMCs, as well as GMCs without star-forming regions.
Among the GMCs associated with star-forming regions, we can see a wide range of star-forming activities:
some GMCs show elevated SFRs, whereas others only host calm star-formation activity.

\subsection{Radial distributions of molecular and atomic gas and SFR}

\begin{figure*}
  \begin{center}
    \FigureFile(80mm,80mm){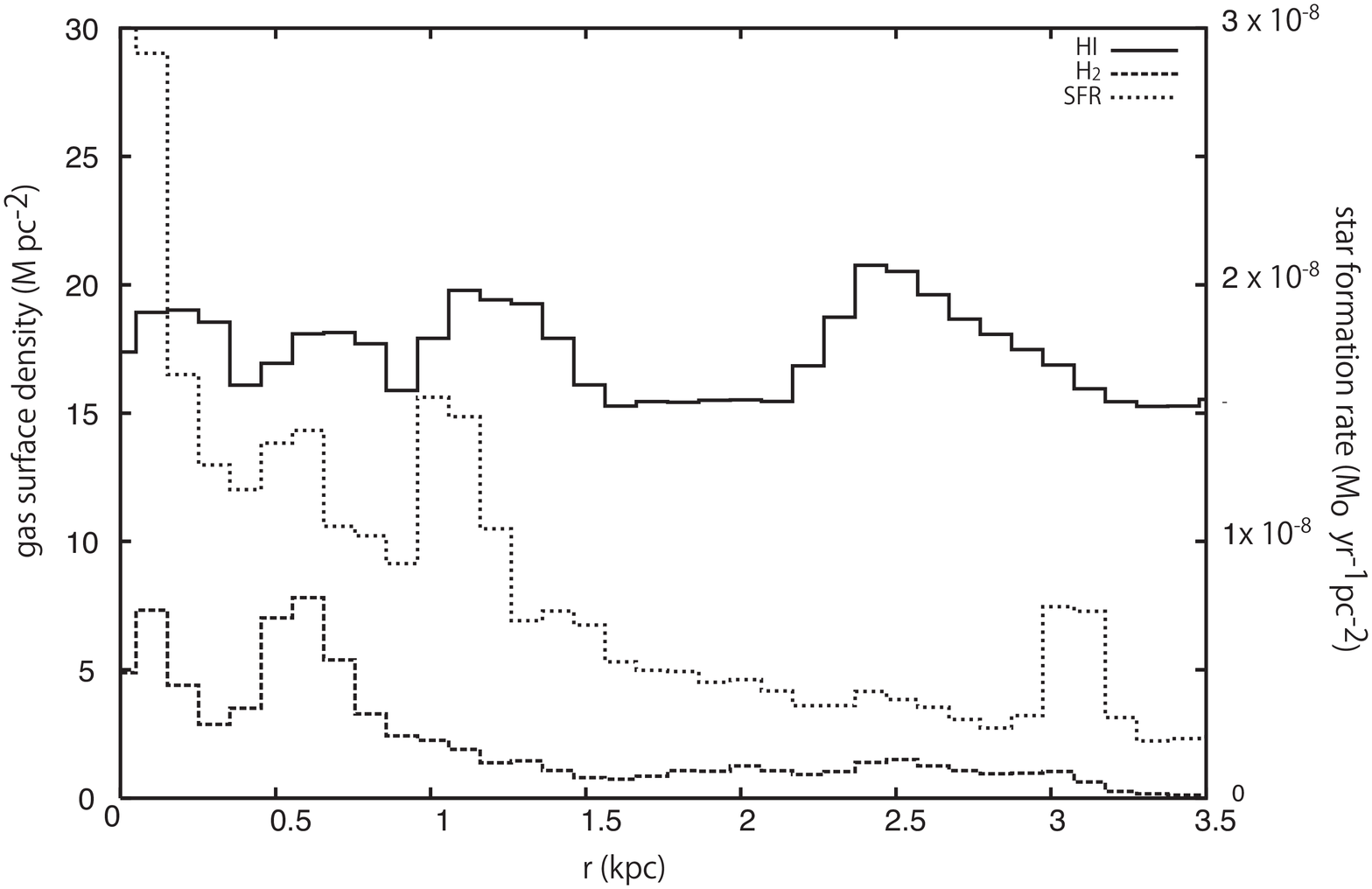}
  \end{center}
  \caption{Radial distribution of the surface densities of H{\sc i} (solid line), H$_2$ (dashed line), and star formation rate (SFR; dotted line).}\label{fig:4}
\end{figure*}

Radial distributions of the molecular and atomic gas surface densities
together with SFR are displayed in figure~\ref{fig:4}.
They are de-projected considering the inclination of M33.

We found that the surface density of atomic gas is higher than that of molecular gas at all radii in this figure. 
This tendency is slightly different from the previous result obtained by Gratier et al. (2010); 
the molecular gas surface density at central kilopersec is
higher than that of the atomic gas in their result, although overall
tendency on H$_2$ and H{\sc i}  radial distributions are similar to us. 
One of the possible causes for this discrepancy can be due to the lower atomic gas surface density estimated
based on the only VLA data without the correction of missing flux.
The radial distribution of molecular gas surface density shows an overall decline, 
with two significant local peaks at the radii of 0.1 and 0.5 kpc.
By contrast, the surface density of atomic gas is rather flat, although it shows several small peaks.
The radial distribution of SFR is similar to that of molecular gas; again, it shows a decline from the center to the outer part of the galaxy,
and we also found four local peaks in the radial distribution of SFR, at radii of 0, 0.5, 1, and 3 kpc from the center.
The first two peaks near the center correspond to the local peaks in molecular and atomic gas surface densities.
The third SFR peak at $r \sim$ 1 kpc is associated with a H{\sc i} peak, but there is no corresponding CO peak.

\section{Discussion}

\begin{figure*}
  \begin{center}
    \FigureFile(150mm,80mm){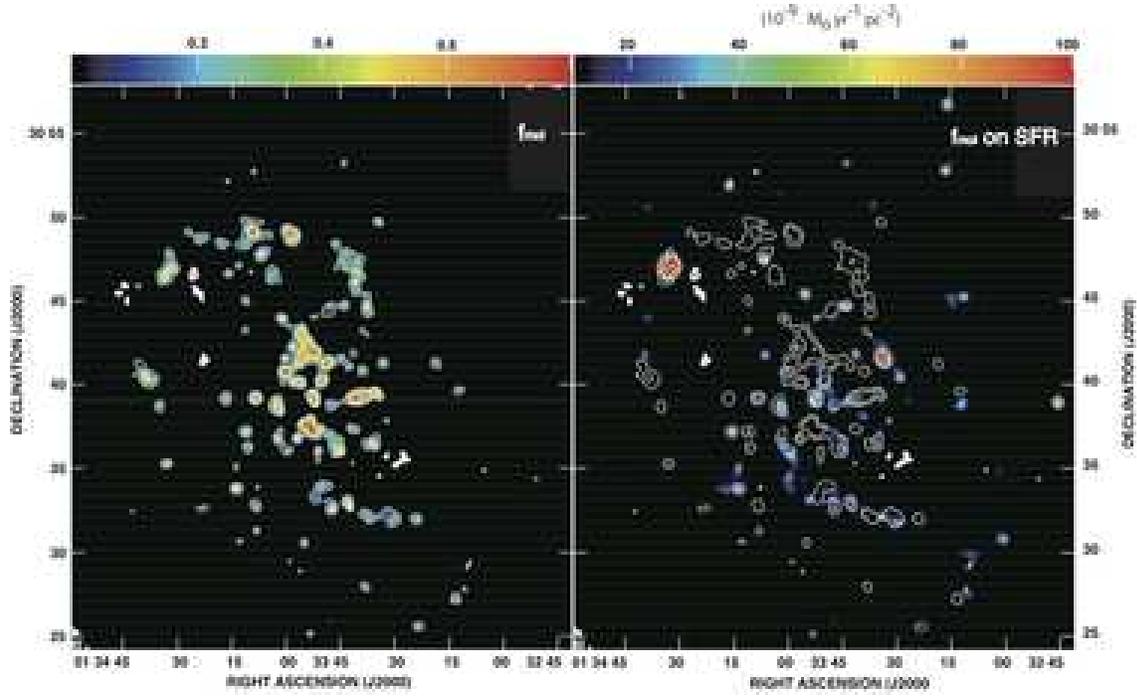}
  \end{center}
  \caption{Molecular fraction ($f_{\rm mol}$) is shown with contours. Contour levels are 0.2, 0.4, and 0.6.
  Colors are $f_{\rm mol}$ in the left  panel and SFR in the right panel.}\label{fig:5}
\end{figure*}

\begin{figure*}
  \begin{center}
    \FigureFile(80mm,80mm){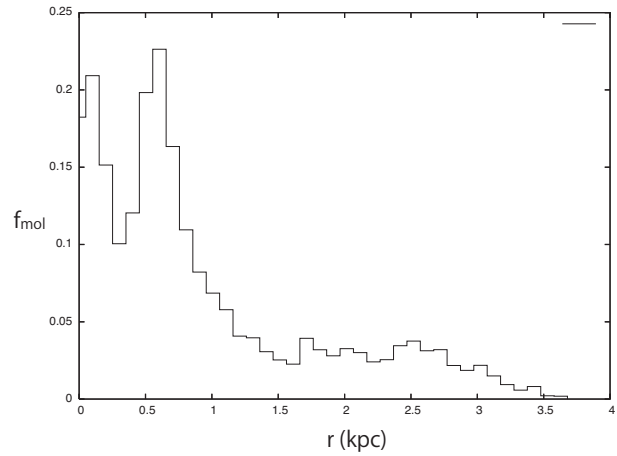}
  \end{center}
  \caption{Radial distribution of $f_{\rm mol}$.}\label{fig:6}
\end{figure*}

As mentioned above (Section 3.2), most of the molecular clouds are associated with local H{\sc i} peaks in the outer regions,
whereas no such clear correspondence can be found in the inner regions.
This suggests that the process of transition from atomic gas to molecular gas is different in the outer and inner regions and that
the process is more efficient in the inner region.
To address the issue quantitatively, we derive the fraction of molecular gas to the total gas surface densities (H{\sc i} + H$_2$), $f_{\rm mol}$,
which is expressed as follows (\cite{elm1993,nak2006a}).

\begin{equation}
f_{\rm mol}=\frac{\Sigma_{{\rm H}_2}}{\Sigma_{\rm HI}+\Sigma_{{\rm H}_2}},
\end{equation}

where $\Sigma_{\rm H_2}$ and $\Sigma_{\rm HI}$ are the molecular and atomic gas surface densities, respectively.
Figure~\ref{fig:5} shows the distribution of $f_{\rm mol}$ values in the panel on the left.
This is the first $f_{\rm mol}$ map covering an entire galaxy with a GMC-scale resolution ($\sim$100 pc).
We found that the $f_{\rm mol}$ values range from $\leq$0.1 to 0.6, showing difference among GMCs.
This figure also reveals that $f_{\rm mol}$ is higher in the inner region than the outer region.
The radial distribution of $f_{\rm mol}$ (figure~\ref{fig:6}) also confirms this finding: it has two peaks near the center and at a radius of $\sim$0.5 kpc,
and decreases toward the outer regions.
This suggests that molecular gas is formed more efficiently in the inner region.
A sharp decrease in $f_{\rm mol}$ is clearly seen around $r = 1\mbox{--}1.5$ kpc, and we suggest that
the molecular front (Sofue et al. 1995; Honma et al. 1995) exists in this region.

The map of $f_{\rm mol}$ is superposed on the SFR map in the panel on the right in figure~\ref{fig:5}.
At the GMC scale, no correlation between H$_2$ and SFR is seen due to
differences in the evolutionary stages of the GMCs (\cite{ono2010}),
and this figure also indicates no correlation between $f_{\rm mol}$ and SFR.
It indicates that the fraction of molecular gas in each GMC is not affected by the current star forming activity.

$f_{\rm mol}$ is determined by ISM pressure ($P$), metallicity ($Z$), and the UV radiation field ($U$) (\cite{elm1993,nak2006a,hon1995, kuno1995}).
Here, in hydrostatic equilibrium for gas and star disk,  total midplane gas pressure is given as follows (\cite{elm1989});

\begin{equation}
P=\frac{\pi}{2} G \Sigma_{gas} (\Sigma_{gas}+ \Sigma_{star} \frac{c_g}{c_s}).
\end{equation}
 
$c_g$ and $c_s$ are velocity dispersions of gas and stars, respectively.
Assuming that the distributions of stars and gas are similar, $\Sigma_{star}$ is proportional to $\Sigma_{gas}$.
Then, $P$ is proportional to the square of the gas surface density, $\Sigma_{{\rm HI+H}_2}$.
In other words, $f_{\rm mol}$ can be expressed as a function of $\Sigma$ for given $Z$ and $U$.

\begin{figure*}
  \begin{center}
    \FigureFile(150mm,80mm){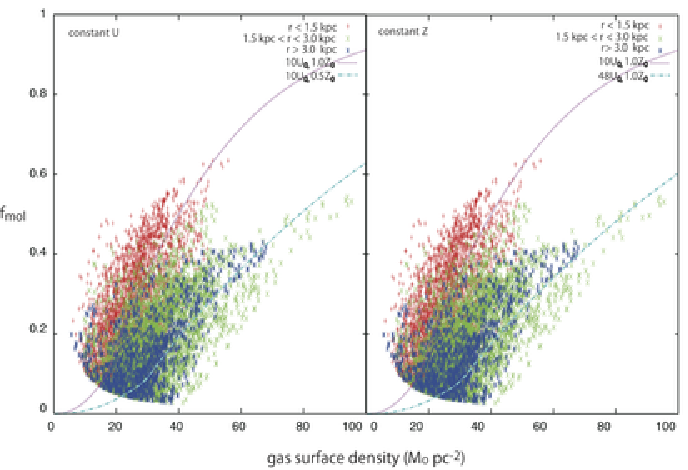}
  \end{center}
  \caption{Correlation between gas surface density and $f_{\rm mol}$.
  Red, blue, and green points represent data with ranges of $r \leq$ 1.5 kpc,  1.5 kpc $\leq r  \leq$ 3 kpc, and  $r \geq $ 3 kpc.
  Pink dotted and  cyan dashed lines indicate correlation for $U=10 U_0$ at $Z = Z_0$ and $Z=0.5Z_0$ (left), and
  for $Z= 1.0 Z_0$ at $ U = 10U_0$ and $U=48Z_0 $ (right), respectively.}\label{fig:7}
\end{figure*}

Figure~\ref{fig:7} shows the correlation between $\Sigma_{{\rm HI+H}_2}$ and $f_{\rm mol}$.
The red, blue, and green dots indicate the measured data points at the inner ($r\leq$ 1.5 kpc), intermediate (1.5 kpc $\leq r \leq$ 3.0 kpc),
and outer radii ($r \geq$ 3.0 kpc), respectively.

We find the presence of two sequences in the $\Sigma_{{\rm HI+H}_2}$--$f_{\rm mol}$ correlation in this figure:
One is the sequence that can be seen in the upper left part of the figure or in the inner region ($r \leq 1.5$ kpc),
showing higher $f_{\rm mol}$ values with respect to a certain $\Sigma_{{\rm HI+H}_2}$.
The other sequence can be found at the lower right part of the figure, i.e., in the outer region ($r \geq 3$ kpc) of the galaxy,
where lower $f_{\rm mol}$ values can be seen compared with the previous $\Sigma_{{\rm HI+H}_2}$--$f_{\rm mol}$ sequence.
This means that we find two distinct $\Sigma_{{\rm HI+H}_2}$ values for a certain $f_{\rm mol}$ value: i.e.,
the $\Sigma_{{\rm HI+H}_2}$ for a given $f_{\rm mol}$ value is higher in the outer part of the galaxy than in the inner part for the same $f_{\rm mol}$.
In other words, for a given gas surface density, $f_{\rm mol}$ value  is higher in the inner part than in the outer part.  
This causes high $f_{\rm mol}$ of GMCs in the inner region shown in figure~\ref{fig:6}. 

Here we investigate the possible causes for the existence of two sequences in the $\Sigma_{{\rm HI+H}_2}\mbox{--}f_{\rm mol}$ diagram.
The first possibility is that molecular gas is formed more efficiently in the inner part of the galaxy. 
Another possibility is that we are overestimating the gas density  in the outer part of the galaxy.
 Dynamical effects such as shock compression due to spiral arms can also be a possible cause. 
These  three  possibilities are examined below.

\subsection{Efficient molecular gas formation -- metallicity and radiation field}

Here we discuss the first possibility.
As mentioned above, $f_{\rm mol}$ can be governed by the interstellar radiation field strength $U$ and metallicity $Z$.
 If the radiation field strength $U$ is constant over the whole radius,
the observed higher $f_{\rm mol}$ value in the inner part of the galaxy can be explained by higher metallicity $Z$.
For instance, the two observed sequences correspond to two distinct metallicities,
i.e., $Z=0.3Z_0$ and $Z=0.15Z_0$, for the inner and outer parts, respectively, at $U=U_0$. Similarly, $Z=Z_0$ and $Z=0.5Z_0$ at $U=10U_0$.
Here, $U_0$ and $Z_0$ are the values of $U$ and $Z$ in the solar neighborhood, respectively.
In short, a difference in $Z$ of a factor of two can explain the two observed sequences if $U$ is constant throughout the galaxy.

In fact, an early observational study on the metallicity distribution in
M33 (\cite{vil1988}) suggests that there is a sharp increase in metallicity in
the central $<$2.4 kpc region. Although the existence of a metallicity gradient within M33 is
still under debate (\cite{ros2008}), we suggest that the
existence of a difference in $Z$ between the inner and outer parts of M33 is a plausible
explanation for the presence of two sequences in the $\Sigma_{{\rm HI+H}_2} - f_{\rm mol}$ diagram.

 On the other hand, a difference in $U$ between the inner and outer parts is needed to explain the two sequences
if we assume that $Z$ is constant across the galaxy.
For example, we suggest that $U=10U_0$ and $U=50U_0$ for the inner and outer parts, respectively, if $Z=Z_0$ at all radii.

Is there any observational evidence for enhanced radiation field
strength in the outer parts of M33?
If the interstellar radiation field originates from massive stars,
elevated star formation should be expected in the outer parts, but
in reality, we see no such tendency, as shown in figure~\ref{fig:4}.
Moreover, if this was the case, we could expect a smaller $f_{\rm mol}$ value for larger
SFR, but again no such correlation is found (right panel in figure~\ref{fig:5}).
It is possible for the strength of the interstellar radiation field to be determined
by intermediate- and/or low-mass stars (e.g., \cite{kom2011}),
but again the strength of $U$ does not account for the observed tendency of $f_{\rm mol}$;
a significant enhancement of $U$ by a factor of 5 is needed in the outer disk,
which is inconsistent with the stellar distribution traced by K-band images.
Considering these facts, we suggest that differences in the radiation
field strength is not the cause for the presence of two sequences
in the $\Sigma_{{\rm HI+H}_2} - f_{\rm mol}$ diagram.
For the first possibility, therefore, the two observed distinct sequences can be explained
by differences in $Z$ between the inner and outer parts.

\subsection{Overestimation of gas density}

Next, we discuss the possibility that we are overestimating $\Sigma_{{\rm HI+H}_2}$ in the outer part.
In figure~\ref{fig:7}, the value of the horizontal axis corresponds to the gas pressure, or volume density, $n$.
$n$ is given by the surface density $\Sigma_{{\rm HI+H}_2}$ multiplied by the scale height $h$, and here we assume that $h$ is constant for all radii.
Therefore, if there is a difference in $h$ between the inner and outer parts of the galaxy, $n$ values should be modified accordingly.
For instance, if $h$ is higher by a factor of 2 at the outer radii,
$n$ should be lower by the same factor.
In this case, the value of the horizontal axis in figure~\ref{fig:7} is overestimated by a factor of 2
and the true data points should be shifted to the left.
If this is the case, the true $\Sigma_{{\rm HI+H}_2}\mbox{--}f_{\rm mol}$ correlation (after correction for overestimation
caused by the differences if $h$) will show just a single sequence.

Is there any observational support for differences in $h$ in M33?
Although no direct measurement of $h$ has been conducted for in M33, it has been pointed out that $h$ is larger in the outer parts than in the inner region
by a factor of two or more in the Milky Way (\cite{Nak2006b}). Hence, we cannot reject this scenario
as a possible cause of the two observed sequences.

It is worth noting that the variation in either metallicity or scale height is rather bimodal---not continuous.
NIR observations have revealed the existence of an $r^{1/4}$ stellar distribution (or bulge)
with a scale length of $8'$ or 2 kpc (\cite{reg1994}).
The putative bulge size of 2 kpc is close to the threshold of these two sequences ($\sim$1.5 kpc),
leading to a speculation that the different environments in and outside the bulge are the cause of the bimodality of the correlation.

\subsection{Dynamical effect}

Another possibility is a dynamical effect such as a shock compression of ISM due to spiral arms.
For example, spiral density wave shocks may push the transition from warm neutral material to cold neutral atomic gas, 
or simply increased self-shielding to allow for the the formation of molecules. Such perturbations may be weaker 
in the outer parts of galaxies owning to the decrease in the stellar density and shallower spiral potentials.
We suggest that this is indeed another possibility, although it may not be a major driver of the observed two sequences, 
because the strength of the spiral density wave in M33 is very weak as mentioned in previous section.

In summary, we have three plausible explanations for the two observed sequences in the $\Sigma_{{\rm HI+H}_2}$--$f_{\rm mol}$ diagram.
The first is the enhanced molecular formation caused by higher metallicity in the inner parts of M33.
Another is the idea that two sequences are observed because of
differences in the scale height of the gas disk between the inner and outer parts of M33,
and the true $\Sigma_{{\rm HI+H}_2}\mbox{--}f_{\rm mol}$ diagram (after correction for the overestimation of $n$)
will show just a single correlation sequence. 
Dynamical effects such as compression of ISM due to spiral arms may also a possible cause.

\section{Summary}

We performed all-disk mapping of $^{12}$CO ($J=1\mbox{--}0$) in the nearby
spiral galaxy M33 using the NRO 45-m telescope.
The observed area covers the central $30^\prime \times 30^\prime$
(7.3 kpc $\times$ 7.3 kpc) of M33, including most of the famous H{\sc ii}
regions such as NGC 604 and NGC 595.
The spatial resolution of the resultant map is 19$^{\prime\prime}$.3,
corresponding to 81 pc, which is sufficient to identify each GMC
in the disk.

Our conclusions are summarized as follows.

\begin{enumerate}

\item The total integrated intensity map of $^{12}$CO ($J=1\mbox{--}0$) emission
shows a clumpy structure with a typical spatial scale of $\sim$100 pc,
corresponding to GMCs. We find no diffuse, smoothly distributed components
of molecular gas at this sensitivity.

\item The overall distribution of molecular gas roughly agrees with
that of H{\sc i}; that is, CO emission coincides with the peaks
of H{\sc i} emission in general. However, close inspection of the CO and H{\sc i} maps
suggests that not every CO emission is associated with local H{\sc i} peaks,
particularly in the inner portions of the disk ($r <$ 2 kpc), although
most of the CO emission is located at the local H{\sc i} peaks in the outer
parts.

\item We find that most of the uncovered GMCs are accompanied by
massive star-forming regions, although the SFRs
vary widely from cloud to cloud.

\item The azimuthally averaged H{\sc i} gas surface density exhibits a flat
radial distribution. However, the CO radial distribution shows
a significant enhancement within the central 1--2 kpc region,
which is very similar to that of the SFR. We find that H{\sc i} is the dominant
component compared with CO at all radii.

\item We obtained a map of molecular fraction, $f_{\rm mol} = \frac{\Sigma_{\rm H_2}}{\Sigma_{\rm HI}+\Sigma_{\rm H_2}}$, at a 100-pc resolution. 
This is the first such map covering an entire galaxy
at GMC-scale resolution. 
We find that $f_{\rm mol}$ tends to be high in the inner region.

\item The correlation between $f_{\rm mol}$ and gas surface density
was investigated. We found two distinct sequences in the
correlation. The presence of two correlation sequences
can be explained by differences in metallicity, i.e., higher metallicity
in the central region ($r<$ 1.5 kpc) than in the outer
parts (a factor of $\sim$2). Alternatively, differences in
scale height can also account for the two observed sequences, i.e.,
increased scale height toward the outer disk.
Dynamical effects such as spiral shocks may also work.

\end{enumerate}


\bigskip

We thank the referee, M. Heyer for his careful review; it significantly improved the manuscript.
We gratefully acknowledge the contributions of the Nobeyama Radio Observatory (NRO) staff to the development and operation of the telescope.
We thank Rene Walterbos and E. Rosolowsky for providing us with the H$\alpha$ and H{\sc i} images of M33, respectively.
TT was financially supported by  JSPS Grant-in-Aid for Scientific Research (C) No. 22540249.
This work is based on observations conducted at the NRO,
which is a branch of the National Astronomical Observatory of Japan, National Institutes of Natural Sciences.





\end{document}